# Context Adaptivity as Enabler for Meaningful Pervasive Advertising

Christine Bauer

Institute for Management Information Systems, Vienna University of Economics and Business, Augasse 2-6, UZA II, 1090 Wien, Austria
chris.bauer@wu.ac.at

**Abstract**. Socio-demographic user profiles are currently regarded as the most convenient base for successful personalized advertising. However, signs point to the dormant power of context recognition. While technologies that can sense the environment are increasingly advanced, questions such as what to sense and how to adapt to a consumer's context are largely unanswered. Research in the field is scattered and frequently prototype-driven. What the community lacks is a thorough methodology to provide the basis for any context-adaptive system: conceptualizing context. This position paper describes our current research of conceptualizing context for pervasive advertising. It summarizes findings from literature analysis and proposes a methodology for context conceptualization, which is currently work-in-progress.

## Introduction

'Pervasive Advertising' promises the possibility to reach out to customers electronically, at any time and anywhere in physical space. The ultimate goal is to influence purchase decisions at the right moment and in an efficient way.

Still, reaching out to the customer in the right spot may not be enough. In the recent years, advertising effectiveness has suffered dramatically. Consumers have become blind to promotional messages as they are overwhelmed by their quantity. Only personalization mechanisms promise to break through the information clutter.

Personalization is a construct from marketing. Based on socio-demographic customer-segmentation and market basket analysis, products, services or content are tailored to consumer needs (Mulvenna et al. 2000). However, thought leaders start to point to the limits of these methods, because the resulting segments are too broad to be effective. Despite socio-demographic clustering, each segment still comprises a heterogeneous group of people. Every person has different needs in distinct situations, what is not considered by socio-demographic segmentation.

Here, the power of context recognition for personalization may remedy the situation, because a wide variety of situational variables come into play. Specific situational user needs can be derived from context and considered for personalized services such as contextual advertising or product recommendations.

Websites (e.g., Linden et al. 2003; Adomavicius et al. 2005; Smith 2004) and mobile (e.g., Yuan and Tsao 2003) personalization mechanisms make use of contextual information already. Technological advancements on the Internet, such as Web crawling and mining, made it possible to easily gather user information. Yet, attempts for transferring context-adaptive advertising to the 'physical world' – such as adaptive digital display advertising (often coined 'digital signage') (e.g., Müller et al. 2009; Goldmedia 2009) or ambient shopping environments (e.g., Maass and Janzen 2007) – are still scarce.

## Experiences gathered

Only recently, some retailers started to integrate technology in their promotional portfolio and upgraded to digital displays ('digital signage'). First experiences with this new medium seem promising: Retailers such as the British supermarket chain 'Tesco' or 'Spar' in Germany increased sales between 25-60% by using digital point-of-sale advertisements (NEC Display Solutions 2006; Page 2007).

The present situation, though, is characterized by business realities, technical hurdles, and scattered research. For instance, advertising agencies still need to adapt their processes and business models to the new opportunities. Currently they seem to apply their 'old' models to the new situations. Additionally, technical challenges are responsible for the small number of more advanced digital signage use, particularly with respect to context- adaptive systems. While research presents grand visions for context computing at large (e.g., Ferscha et al. 2009a; Ferscha et al. 2009b; Black et al. 2009), advancements on smaller scales are scattered. In the field, the research community works on individual problems that need to be solved in different phases of system development (e.g., data collection challenges, data transformation, adaptivity, presentation, etc.). However, the community lacks a coherent big picture, integrating the diverse research threads within the field. As a result, no holistic and systematic methodology has outlined how a context-adaptive system, such as a context-adaptive advertising system, should actually be developed and constructed.

In short, it is not clear yet, how to effectively use digital signage. "*It is neither a static poster, nor is it broadcasting*", as an expert of Smart TV-Networks – one of the Europe's pioneers in digital signage – points out (Goldberg 2007). "*It is a hybrid – narrow-casting –, as we can selectively reach a small target group with moving images. And we still do not know how to challenge this*."

## Results obtained

The majority of researchers in the field (who frequently are computer scientists) tends to concentrate on system architectures, prototypes and toolkits (Baldauf et al. 2007; Dey and Abowd 2000; Hong et al. 2009) or data capture and aggregation challenges (Ferscha et al. 2002). By taking this viewpoint, they neglect that context-adaptive applications are part of a socio-technical system, which comprises as well a human perspective.

Yet, although concentrating on technical issues, research in the field of context-adaptivity is scattered: It is often unclear whether 'context-aware' or 'context-sensitive' systems are the same as 'context-adaptive' ones. Researchers in the field seem to give different names to similar problems.

Little systematization of these diverse activities has occurred. As real-world deployments emerge, a more structured view of the field's activities and achievements may be beneficial. Meeting this challenge, we propose a process model for context adaptivity that integrates the different research threads of context-aware computing. This model also provides an overview of the sequence of challenges engineers face when designing a fully functional and meaningful adaptive service such as contextual advertising.

As depicted in Fig. 1, there are four phases of challenge (Bauer and Spiekermann 2011).

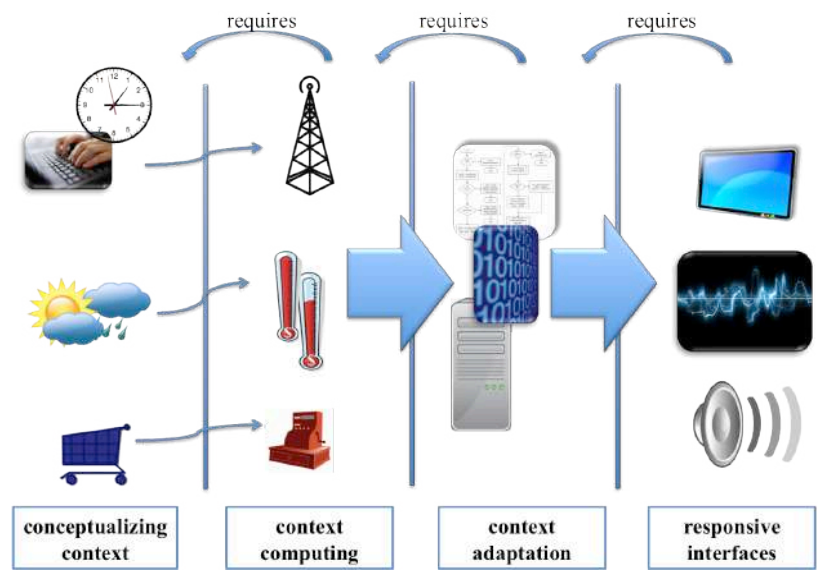

Fig. 1 The process of context adaptivity (Bauer and Spiekermann 2011)

The first step is to 'conceptualize context': to systematically identify the full spectrum of context variables that can be used to meaningfully interpret a specific adaptive service (Bradley and Dunlop 2005; Dourish 2004). For example, a context model for selling hedonistic products could foresee temperature and number of persons as relevant context. Other variables could be, for example, hair color, content of the shopping basket, time of the day or consumer profile.

In the second step of the process, 'context computing', relevant sources of context information are identified and collected (Dey 2001). For identifying and collecting relevant data sources, we use sensors and other contextual sensing technologies (Pascoe 1998) such as radio-frequency identification (RFID) or eye-tracking. In terms of the example this means, for instance: A temperature sensor measures 30°C; an eye-tracking system counts two persons staring at a certain display. Temperature could also be obtained from a weather forecast website and number of persons could be counted with a video analysis. The stored context information is used to trigger certain events (Ferscha et al. 2004).

Such a trigger marks the beginning of the 'context adaption' stage of the process. The goal of this phase is to intelligently adapt to the context that has been detected. Adaptivity mechanisms then use an algorithm to translate the captured context into the desired action. In terms of the example this could mean: The system computes for the context set 'two persons, 30°C' to show ice cream advertising with addressing consumers in plural form. Other computed outcome could be displaying the temperature or choosing warm colors.

Finally, in the responsive interfaces stage, the computed personalization action is operationalized and presented. In terms of the example this could be operationalized as a written message in bold letters on a display: "*It is hot in here. What about a dish of ice cream for both of you?*" Other thinkable presentations could be, for example, displaying a picture of ice cream or spreading fragrances, which could give an appetite for ice cream.

The remainder of this paper will focus on the first phase of context adaptivity: conceptualizing context.

## New ideas and ongoing research

"*How are dimensions of context identified, quantified, and interrelated for each situational purpose?*" (Bradley and Dunlop 2005). The first step to answering this question is to conceptualize context (see also Fig. 1).

> We define context conceptualization as ***the process*** by which a personalization situation is deconstructed into measurable and logically disjunctive information units, all of which must be combined to create an adaptive service.

Some scholars already worked on this issue by identifying disjunctive information units. Common information categories include a user's location and environment, the identities of nearby people and objects (entities), and changes to those entities (Dey 1998; Schilit and Theimer 1994). These concepts are, though, very generic and convey broadest sense only. As a result, these cannot be directly operationalized for constructing an adaptive system.

Schmidt et al. (1999) elaborate a more systematized approach for context-aware mobile systems. They relate context information to two domains: human factors and the physical environment, both in the broadest sense. They then deconstruct human factors into information about users themselves, their social environment and their tasks. Likewise, the physical environment is described by location, infrastructure, and physical conditions.

Tarasewich (2003) differentiates three categories: environment, participants, and activities. Representative characteristics of environment are location and orientation of objects, physical properties, brightness and noise levels, as well as availability and quality. Participants are characterized by their location and orientation, personal properties, mental state, physical health, and expectations. Activities are described by tasks and goals of participants as well as events in the environment such as weather.

For all three categories the model considers time with respect to present, past, and future.

Bradley and Dunlop (2005) take a multi-disciplinary approach to context, integrating linguistics, psychology, and computer science. Combining and building upon existing models from these three domains, they distinguish the world of the 'user' from that of the 'application'; for both, they consider 'incidental' and 'meaningful' context. In a circular layer that surrounds both worlds, they add a third 'contextual' world, which they break down into six dimensions: task, physical, social, temporal, cognitive and application's context.

However, these models are still too generic to be operationalized for an adaptive application. Furthermore, these models take an approach with predefined categories, which appears insufficient for such a dynamic phenomenon like context. It is false hope to be able to come up with a comprehensive and complete list (or model) of context categories that fits several purposes. Instead, we argue that it is only possible to understand a context situation and identify available and meaningful information for adaptive service delivery by means of a process methodology.

Essentially, there are two problems. The first is creating a context model; the second is identifying the information required in a concrete system. Conceptualizing context means identifying the information units required. Having a concrete model can help identifying the information.

Conceptualizing context for pervasive advertising involves the consideration of the advertiser, the brand, the company's objectives and the advertising campaigns. These variables are domain-specific and are hardly covered by existing context taxonomies. Other possible considerations for pervasive adverting include highly dynamic issues. For instance, the stage of the buying process, in which consumers find themselves at the moment when an advertisement is displayed, could be considered. Other dimensions are a consumer's emotional state or behavior, the noise in the environment or the odor. Yet, none of these variables are considered by existing context models and information taxonomies.

Against this background, we suggest one process methodology that can be applied to conceptualize the context for a pervasive context-adaptive advertising service. This approach may be reused for other application domains and can amend and replace existing information categories. Additionally, we want to stress that conceptualizing context should not be (purely) technology-driven; instead, the identification of information units should be driven by business needs.

All traditional system design and development methodologies include a requirements engineering phase (Kurbel 2008), where stakeholder interests are collected. Further, business expectations are formulated and potentially drawn from the strategic goals of a company or system operator. Moreover, the tool-level tasks of users are analyzed (Te'eni et al. 2007).

When engineering requirements for context-adaptive advertising, however, it is difficult to describe people's implicit or indirect interaction with the system. Users often only interact with the system indirectly, while engaged with another primary task. For example, a person who is shopping may perceive an adaptive advertising screen only in the periphery. Advertising may be effective although the consumer has not consciously seen the advertisement As a result, engineers often do not know what data they should collect and how they should combine them in a way that creates a meaningful adaptive service. User-centered designers gather information by interviewing users about their tasks. But what should users be asked in the context of advertising, when they tend to perceive advertising screens only in the periphery? And how can engineers determine what information they should collect and combine to create a meaningful adaptive service if they cannot interview users on their 'non-primary task'?

To overcome these challenges, we suggest an 'early' requirements engineering phase that aims to 'conceptualize' context. As outlined in the beginning of this section, conceptualization of context identifies the full spectrum of measurable and logically disjunctive information units available and/or required for a meaningful service. It deems necessary to recognize the full spectrum of available information units to tap the full potential of an adaptive service. Particularly, in designing pervasive advertising systems, both advertisers and engineers need to agree on what is meant by 'context'. Conceptualizing context helps these parties agree on which components of a context are important.

At our institute, we conceptualized context for pervasive context-adaptive advertising in the retail outlet (Bauer and Spiekermann 2011). Engaging in the conceptualizing work, we presumed to use a networked digital signage system with displays spread out in a retail outlet. These displays are used as interface to transmit highly personalized context-adaptive advertising messages.

Our conceptualization approach is both top-down and bottom-up (Fig. 2). The top-down approach is informed by a literature review and involves reflecting on the overall dimensions of the envisaged system, decomposing it into smaller parts (compositional subsystems). In contrast, the bottom-up approach considers information units in each of the identified dimensions, which are pieced together to form grander systems. We used these two perspectives to analyze the service context in three phases, refining our analysis each time.

More precisely, our first step was to identify relevant literature from the multidisciplinary (human-computer interaction and computer science) domain of context. We evaluated existing approaches and finally started out from the established context model by Schmidt et al. (1999). Taking a top-down approach, we broke down 'context' into six subsystems (consumer profile, social environment, buying process, infrastructure, location, and conditions), in a similar way as Schmidt et al. (1999) did it. Taking a bottom-up approach, we imagined a concrete situation in a shopping mall where a context-adaptive advertising display could be installed. We thought of context variables that might be useful for the system to adapt to; then we pieced the variables together to form grander categories.

In the verification phase, we invited experts to interpret and adapt the model taking a top-down approach. In a second exercise, we asked them to recall one of their last shopping experiences in a physical store and imagine that they encountered a context-adaptive advertisement. They had the opportunity to adapt the model by rearranging the sticky notes or introducing new ones.

Still, this approach for conceptualizing context is work-in-progress and needs further verification phases to verify its applicability for other context-adaptive services of pervasive advertising and also in other domains.

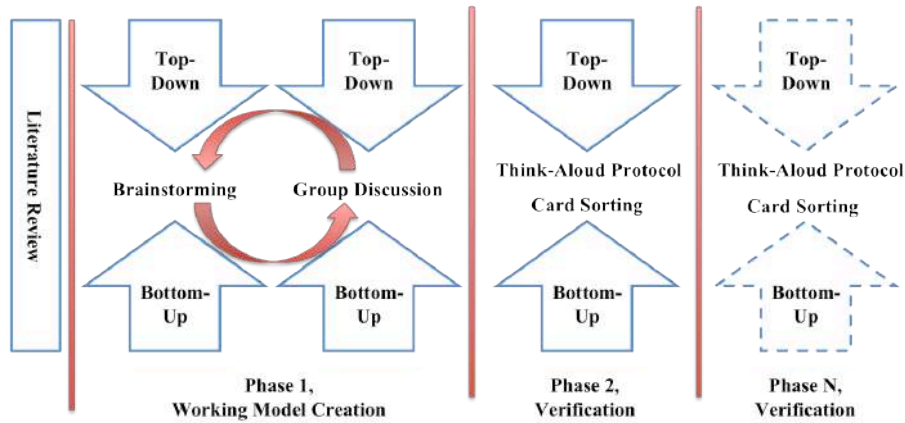

Fig. 2 Methodology for conceptualizing context. The top-level domain is broken into its compositional subsystems (top-down) while individual base elements are pieced together to form grander systems (bottom-up)

## Vision of advertising and shopping in 25 years from now

Our vision is that advertising and shopping will be strongly interweaved. Every product advertisement will offer the opportunity for an immediate purchase – either directly on a public display or via a smart phone, which will connect with the public display via wireless LAN or other technologies.

Still, we think that advertising will be more targeted on creating a 'good' shopping atmosphere or entertainment than on selling individual products. Especially, price competition advertisements will be less seen on public displays than in current days. Particularly in retail premises, large displays will be used to enhance the shopping experience.

We envision that more factors than merely socio-demographics will be considered for personalized advertising. As technology advances and more experience is gained with their effective applications, companies will integrate various context variables to let display advertisement respond to the environment. Specifically, the situational user profile, which is dynamic and may change within seconds (e.g., mood, affect, behavior), will be largely considered by such systems.

## Open questions on topics related to pervasive advertising

**RQ 1.** Which type of context information has to be considered (and captured) by context-adaptive information systems (e.g., location, identity, brain activity, temperature, historic activities such as past purchases, etc.) to serve market players' needs (e.g., raising human attention for advertisements)?

**RQ 2.** In which ways should context-aware advertising adapt to context (i.e., how should customization and personalization be undertaken)?

**RQ 3.** In advertising, does environmental, i.e. impersonal, context (e.g., temperature), person-related context (e.g., number of persons gazing at a public display), or personal on an individual level context (e.g., individually set up personal identity profile) more strongly affect a person's memory (brand recall and recognition)?

**RQ 4.** How can context-adaptive advertising be applied such that advertising is both subtle and effective?